\documentclass[prb,twocolumn,superscriptaddress,showpacs,floatfix,reprint]{revtex4-1}

\usepackage[latin1]{inputenc}
\usepackage{graphicx}
\usepackage{amssymb}
\usepackage{amsmath}
\usepackage{bm}
\usepackage{multirow}

\begin{document}

\title{Violation of the Rule of Parsimony: Mixed Local Moment and Itinerant Fe Magnetism in Fe$_{3}$GeN}

\author{Mari Tsumuraya}
\affiliation{Department of Physics and Astronomy, University of Missouri, Columbia, Missouri 65211 USA}
\author{David J. Singh}
\email{singhdj@missouri.edu}
\affiliation{Department of Physics and Astronomy, University of Missouri, Columbia, Missouri 65211 USA}
\affiliation{Department of Chemistry, University of Missouri, Columbia, Missouri 65211 USA}
\affiliation{Department of Mechanical and Aerospace Engineering, University of Missouri, Columbia, Missouri 65211 USA}

\date{\today}

\begin{abstract}
Ternary iron nitrides are of considerable interest due to their diverse magnetic properties.
We find, based on
first principles calculations,
that the relatively minor structural distortion from the cubic antiperovskite structure
in Fe$_3$GeN,
consisting of octahedral rotations,
leads to unusual magnetic behavior. In particular, there is a separation into Fe sites with very
different magnetic behaviors, specifically a site with Fe atoms having a stable local moment and
a site where the Fe shows characteristics of much more itinerant behavior.
This shows a remarkable flexibility of the Fe magnetic behavior in these nitrides and points
towards the possibility of systems where minor structural and chemical changes can lead to 
dramatic changes in magnetic properties.
The results suggest that, analogously to oxide perovskite materials, modulation of magnetic properties via
chemical or strain control of octahedral rotation may be feasible.
This may then lead to approaches for tuning magnetism to realize properties of interest, for
example tuning magnetic transitions to quantum critical regimes or to proximity to metamagnetic
transitions of interest for devices.
\end{abstract}

\maketitle

\section{Introduction}

Iron nitride compounds and particularly
their magnetic properties have been the subject of considerable interest.
This was stimulated by reports of exceptional magnetic properties in tetragonal
Fe$_{16}$N$_2$, including high magnetization in combination with
reasonable coercivity for many applications.
\cite{Jack1948,takahashi,Wang2020}
Related to this, Fe shows a remarkable variety of magnetic behaviors depending on structure.
For example, the ground state bcc structure of $\alpha$-Fe is a well known ferromagnet,
while austenite, fcc structure shows much weaker magnetism, and lies higher in energy
because of this difference.
\cite{bramfitt,fu}
Other binary Fe-N compounds include Fe$_3$N
which is hexagonal and ferromagnetic, \cite{zieschang}
and Fe$_4$N, which is cubic, with a lattice consisting of Fe on fcc sites, and N
in octahedral holes. This compound, despite its fcc Fe sublattice, has a substantial magnetization of 2.14 $\mu_B$/Fe,
similar to bcc Fe.
\cite{chen,costa-kramer}

Ternary iron-nitride compounds provide opportunities for better understanding the relationships between
structure and magnetism in iron rich phases, but have been relatively less investigated as compared to the
binary phases.
Some examples include
GaFe$_{3}$N, which is an antiferromagnet,
\cite{burghaus}
and Fe$_{3}$RhN,
which is an itinerant ferromagnet with a magnetic moment of 8.3 $\mu_B$ per formula unit.\cite{Houben2006}.
These compounds form in the cubic antiperovskite structure, which is the same as the structure of
Fe$_4$N, with one Fe substituted by another element.
The antiperovskite structure may thus be regarded as fcc Fe with one quarter of the Fe atoms substituted
by another element to yield a simple cubic lattice and N inserted into Fe coordinated octahedral holes.

Other Fe-N ternary phases with this structure include
AlFe$_3$N, \cite{fu-2014}
and ferromagnetic ZnFe$_3$N. \cite{fu-2014,niewa}
Antiperovskite intermetallics, stoichiometry $A$Fe$_3$N, where $A$ is a metal or metalloid,
normally form in the undistorted cubic $AB$O$_3$ perovskite type structure, space group $Pm\bar{3}m$
with $A$ on the perovskite $A$-site, (0,0,0), Fe on the perovskite O-site,
(0,\textonehalf,\textonehalf), (\textonehalf,0,\textonehalf), (\textonehalf,\textonehalf,0),
and N on the perovskite $B$-site (\textonehalf,\textonehalf,\textonehalf).
The perovskite structure is often described as consisting of corner sharing $B$O$_6$ octahedra and $A$-site cations
in the large interstices between these octahedra.
Oxide perovskites most commonly distort from the cubic structure by rotations and tilts of the $B$O$_6$ octahedra.
\cite{glazer}
Antiperovskite intermetallics are distinct in this regard in that they normally form in the ideal cubic structure,
without distortions.

\begin{figure}[tbp]
\centerline{\includegraphics[width=\columnwidth]{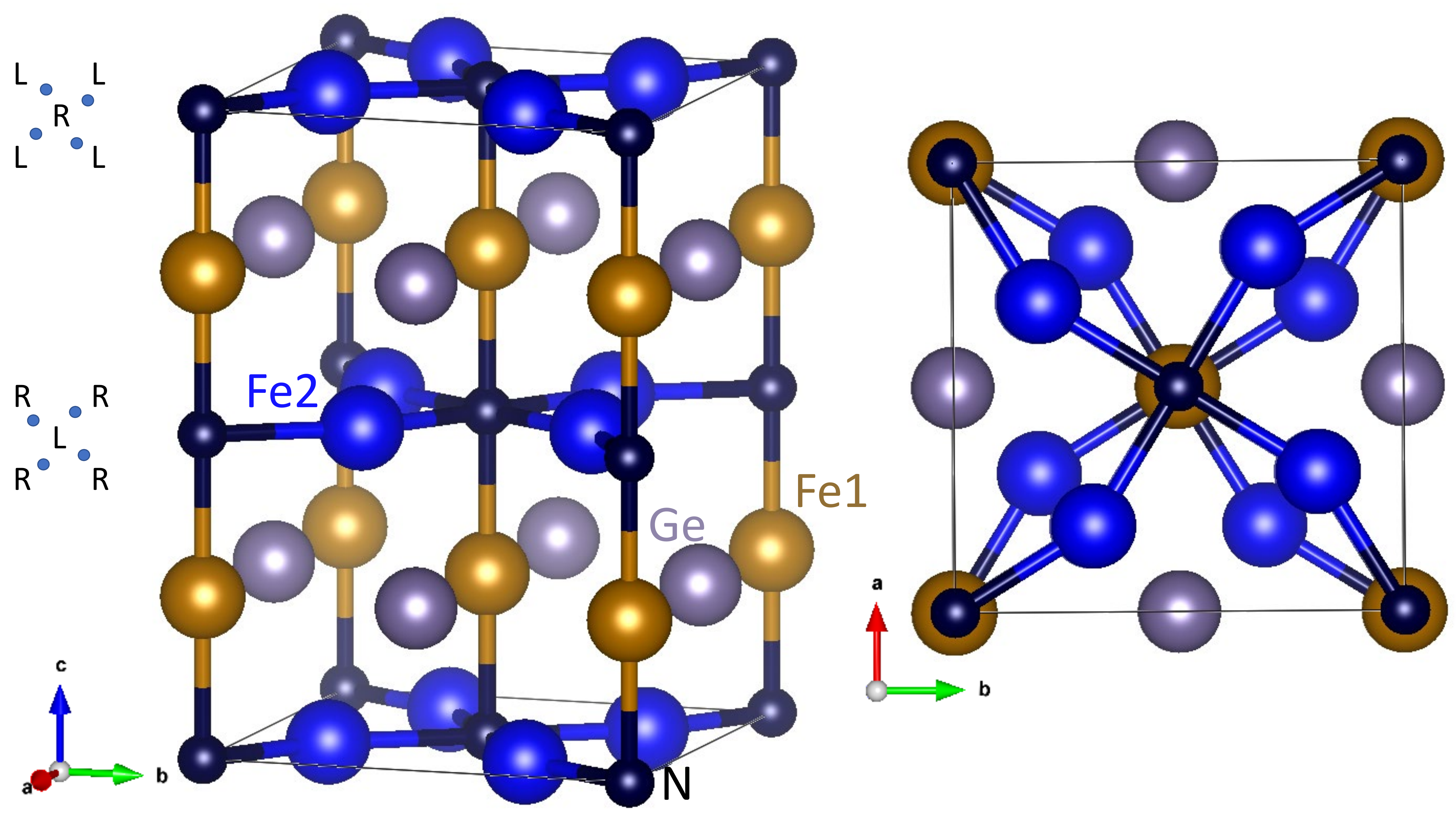}}
\caption{Crystal structure of Fe$_{3}$GeN shown in two views as indicated.
The atomic positions are as optimized by energy minimization using the PBE GGA.
Note the alternating rotation of the NFe$_6$ octahedra around the $c$-axis,
as indicated by the schematics for two layers on the left side of the figure.}
\label{Structure}
\end{figure}

Iron germanium nitride Fe$_{3}$GeN is exceptional in this regard.
In particular, it has an antiperovskite based
tetragonal structure, which consists of corner sharing NFe$_6$ octahedra that are alternately rotated
along the $c$-axis of a tetragonal cell, similar to many oxide perovskites.
Interestingly, while not common among intermetallic antiperovskite compounds,
octahedral rotation and tilts are very important in modifying the properties
of oxide perovskites.
\cite{he,rondinelli,bilc}
Scholz and Dronskowski\cite{Scholz2017} found that the structure of
Fe$_{4-x}$Ge$_x$N$_{y}$ changes from a cubic antiperovskite structure to the
distorted tetragonal structure with increasing content of Ge replacing Fe.
Liu and coworkers showed that this transition is
continuous as a function of Ge content in Fe$_{4-x}$Ge$_x$N.
\cite{Liu2020}
In any case, Fe$_3$GeN,
which shows a net magnetic moment 0.2-0.3 $\mu_B$/Fe with sample dependence,
has tetragonal symmetry at room temperature.
\cite{boller1968,Scholz2017}
This structure has a doubled primitive unit cell containing two formula units,
space group $I4/mcm$, number 140.

Previous studies report some unusual physical behaviors including
an unusual critical behavior.
This is reported to be
intermediate between the expected short range Heisenberg type behavior and more long range
behavior. \cite{Kan2017} In addition, an anomalous Hall effect is reported.\cite{Kan2016}
One complication has been that the compound may form off stoichiometry, in particular with partial
filling of the N-sites.
\cite{boller1968,Scholz2017}
This tendency to variable stoichiometry complicates analysis of the physical properties.
Reported experimental lattice parameters are
$a$=5.231 \AA, $c$=7.658 \AA, for GeN$_{0.5}$Fe$_{3}$,
\cite{boller1968}
and $a$=5.3053 \AA, $c$=7.7632 \AA, for Fe$_{3}$GeN.
\cite{Kan2017}
The difference between them is presumably mainly due to the nitrogen stoichiometry.

In any case, octahedral rotations lead to distinct Fe sites, specifically
the three Fe atoms in the formula unit are divided into two plane Fe atoms and an apical Fe atom,
which may then behave differently with respect to formation of the electronic structure.
These are the Fe1 site (Wycoff site, 4$a$, one atom per formula unit, two per primitive cell)
and the Fe2 site (Wycoff site 8$h$, two atoms per formula unit, four per primitive cell).
Considering the intersecting chains of Fe and N atoms along the Cartesian directions
that are characteristic of the perovskite structure, the octahedral rotation in Fe$_3$GeN
leaves straight ... Fe1-N-Fe1-N ... chains along the $c$-axis direction, while the
... Fe2-N-Fe2-N ... chains in the in-plane directions become bent, specifically with N-Fe-N
bond angles that deviate from 180$^\circ$.
The calculated octahedral rotation angle for the PBE GGA at the experimental lattice parameters is
13.24$^\circ$, leading to a N-Fe2-N bond angle of 153.52$^\circ$.
Prior theoretical work has shown that the ferromagnetic moment observed is a consequence of
a ferrimagnetic ordering of the Fe1 and Fe2 itinerant moments associated with Fe $d$ states around the
Fermi level.
\cite{Liu2020}
The presence of Fe $d$ dominated electronic states near the Fermi level and associated magnetism
is a common feature of Fe-based antiperovskite nitrides.
\cite{Mohn1992,Rebaza2011}
The purpose of the present work is to investigate the effects of the
structural separation into two distinct Fe sites.

\section{Computational Methods}

The present first principles work was done within density functional theory
using
general potential linearized augmented plane-wave (LAPW) method,
\cite{wien2klapw}
as implemented in the WIEN2k code.
\cite{wien2kcode}
The reported results were obtained using LAPW sphere radii of
2.30 Bohr, 1.55 Bohr and 2.00 Bohr for Ge, N and Fe, respectively.
We tested other choices of sphere radii and used somewhat smaller sphere radii of 1.85 Bohr
for Fe in the determination
of the equilibrium lattice parameters in order to avoid sphere overlaps.
The calculations were done with well converged, tested choices for the basis set
cutoff and the Brillouin zone sampling.
The results shown were obtained with Brillouin zone sampling
by 16$\times$16$\times$16 uniform {\bf k}-point meshes.
We did calculations with both the common Perdew, Burke and Ernzerhof generalized gradient approximation
(PBE GGA)
\cite{pbe}
and with the local density approximation (LDA) due to the possible sensitivity of
results on magnetism of an Fe-based intermetallic to the particular functional. \cite{fu}
The differences may be regarded as an indication of uncertainty due to the choice of the functional.

The nature of the magnetism was studied using constrained density functional theory.
\cite{dederichs}
This allows calculations of the energy and other properties, for example, contributions of different sites
to the magnetism, as a function of a constraint. Here we used
the fixed spin moment method, \cite{fsm1,fsm2} where the constraint is on the total spin magnetization of the
unit cell. This constraint is imposed by fixing the difference 
between the spin-up and spin-down occupations. This leads to a difference in Fermi energy between
spin-up and spin-down corresponding to the constraining field.
\cite{wien2klapw}

\begin{table*}[t]
\caption{Ground states from self consistent field calculations. M$_\text{total}$ is the total magnetic moment per
six Fe atom unit cell.
M$_\text{Fe1}$ and M$_\text{Fe2}$ are the Fe1 and Fe2 magnetic moments from integration
of the spin densities in the LAPW spheres, and are given per atom. Note that spin density also has
small components outside the Fe spheres, which accounts for the difference between the total moment per cell
and the sum of the Fe moments.
These are given at the experimental
and at the calculated lattice parameters, $a$ and $c$.
for the LDA and PBE functionals.
Note that the unit cell contains two Fe1 atoms and four Fe2 atoms.}

\centering
\begin{tabular}{|c|cccc|cccc|}
\hline\hline
 \multirow{2}{*}{Functional} &\multicolumn{4}{c|}{Experimental Lattice Parameters} &\multicolumn{4}{c|}{Theoretical Lattice Parameters}\\ \cline{2-5} \cline{6-9}
 \multirow{2}{*}{ } &Lattice &\multicolumn{3}{c|}{Magnetic Moment ($\mu_{B}$)} &Lattice &\multicolumn{3}{c|}{Magnetic Moment ($\mu_{B}$)}\\
 &Parameters (\AA) &M$_{total}$ &M$_\text{Fe1}$ &M$_\text{Fe2}$ &Parameters (\AA) &M$_{total}$ &M$_\text{Fe1}$ &M$_\text{Fe2}$\\
 \hline
 \multirow{2}{*}{PBE} &$a$=5.3053 &\multirow{2}{*}{4.66} &\multirow{2}{*}{1.86} &\multirow{2}{*}{0.31} &$a$=5.2133 &\multirow{2}{*}{4.47} &\multirow{2}{*}{1.72} &\multirow{2}{*}{0.30}\\
 &$c$=7.7632 & & & &$c$=7.7638 & & & \\
\hline
 \multirow{2}{*}{LDA} &$a$=5.3053 &\multirow{2}{*}{4.24}              &\multirow{2}{*}{1.66} &\multirow{2}{*}{0.29} &$a$=5.1072 &\multirow{2}{*}{3.23} &\multirow{2}{*}{1.07} &\multirow{2}{*}{0.30}\\
 &$c$=7.7632 & & & &$$c$$=7.5312 & & & \\
\hline\hline
\end{tabular}
\label{SCF_table}
\end{table*}

We did calculations at the reported experimental lattice
parameters of
$a$=5.3053 \AA, and $c$=7.7632 \AA,
\cite{Kan2017} and also as a function of lattice parameters.
In all cases, we relaxed the free internal coordinates, specifically the in-plane Fe2
position, which is related to the octahedral rotation.
For the PBE functional at the experimental lattice parameters, we obtained an Fe2 position of (0.691,0.191,0).
This corresponds to a larger 
rotation than that for the coordinate
reported by Boller and coworkers, \cite{boller1968} who obtained (0.73,0.23,0)
for samples with a reported stoichiometry Fe$_3$GeN$_{0.51}$. The lower rotation in that report is
intermediate between the unrotated cubic structure of Fe$_3$Ge and our result for stoichiometric Fe$_3$GeN.
Interestingly, this trend is also consistent with the trends for oxide
$AB$O$_3$ perovskite materials.
In those oxides,
increasing the $B$-site size (N here), while keeping the $A$-site fixed, reduces the perovskite
tolerance factor,
$t=(r_A+r_{\rm O})/\sqrt{2}(r_B+r_{\rm O})$, where $r_A$, $r_B$ and $r_{\rm O}$ are the ionic radii of
$A$, $B$ and O, respectively.
\cite{lufaso}

Reduction of the tolerance factor in normal perovskites generally leads to stronger octahedral rotation.
One may also rationalize the result of Scholz and Dronskowski in this picture.
They reported that
Fe$_{4-x}$Ge$_x$N$_y$ changes from cubic to tetragonal as $x$ increases. \cite{Scholz2017}
This is by noting that
Ge is smaller than Fe, based on trends in transition metal germanides, also
consistent with a shrinking unit cell volume as Fe is alloyed with Ge.
Thus the tolerance factor is decreased and the tendency to rotation is increased
when the Fe on the $A$-site is replaced by Ge.
Thus while the bonding of these compounds is covalent and metallic,
\cite{Mohn1992}
and not ionic as in the
oxide perovskites, some of the structural rules, in particular the role of atomic sizes
carry over.

\section{\textbf{Results and Discussion}}

\begin{figure}[tbp]
\includegraphics[width=0.82\columnwidth]{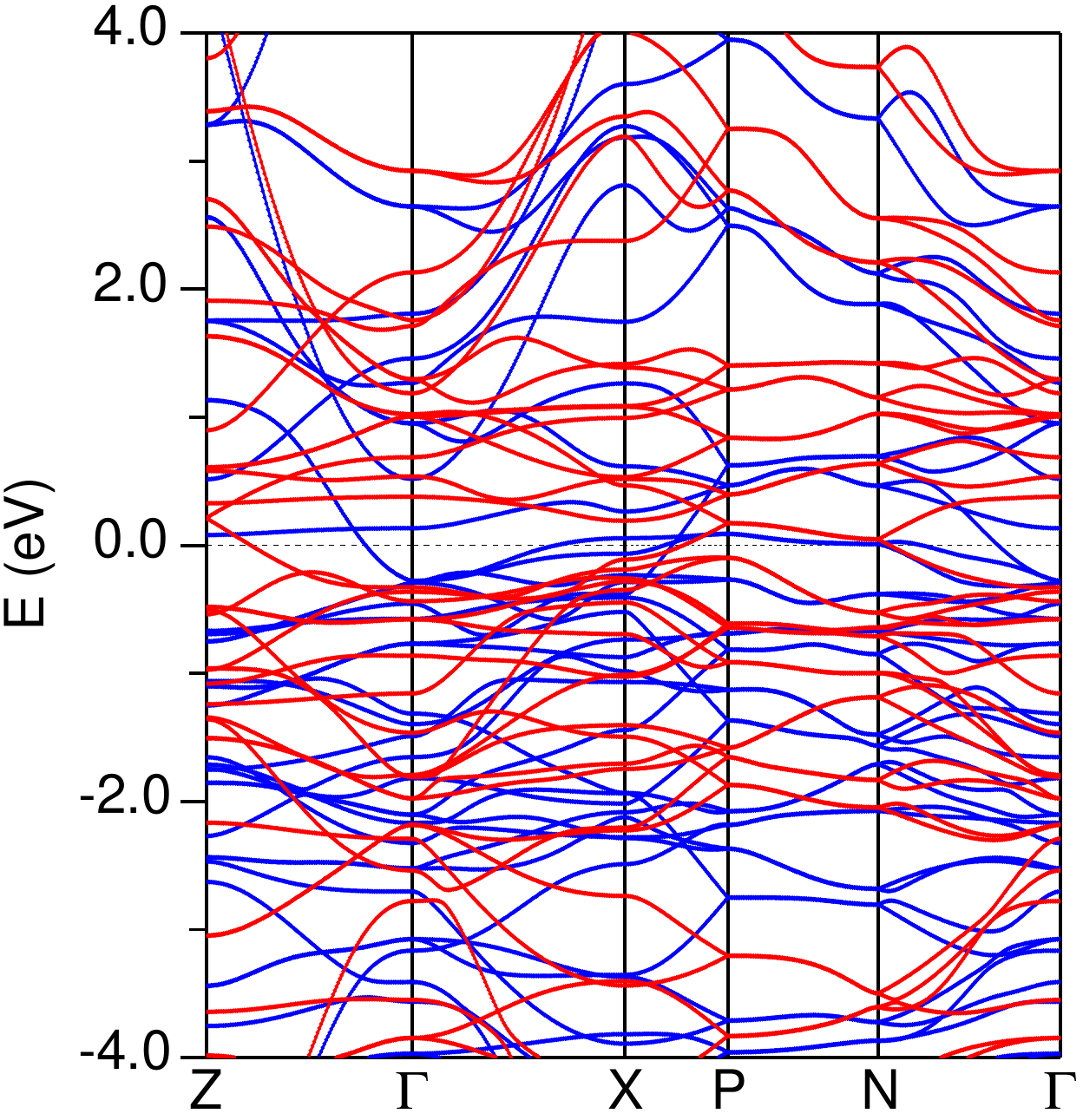}
\caption{Calculated band structure for majority and minority spin. The dashed line denotes the Fermi level at 0 eV.
Majority spin bands are shown in blue and minority spin bands as red.}
\label{fig:band}
\end{figure}

\begin{figure}[tbp]
\centerline {\includegraphics[width=\columnwidth]{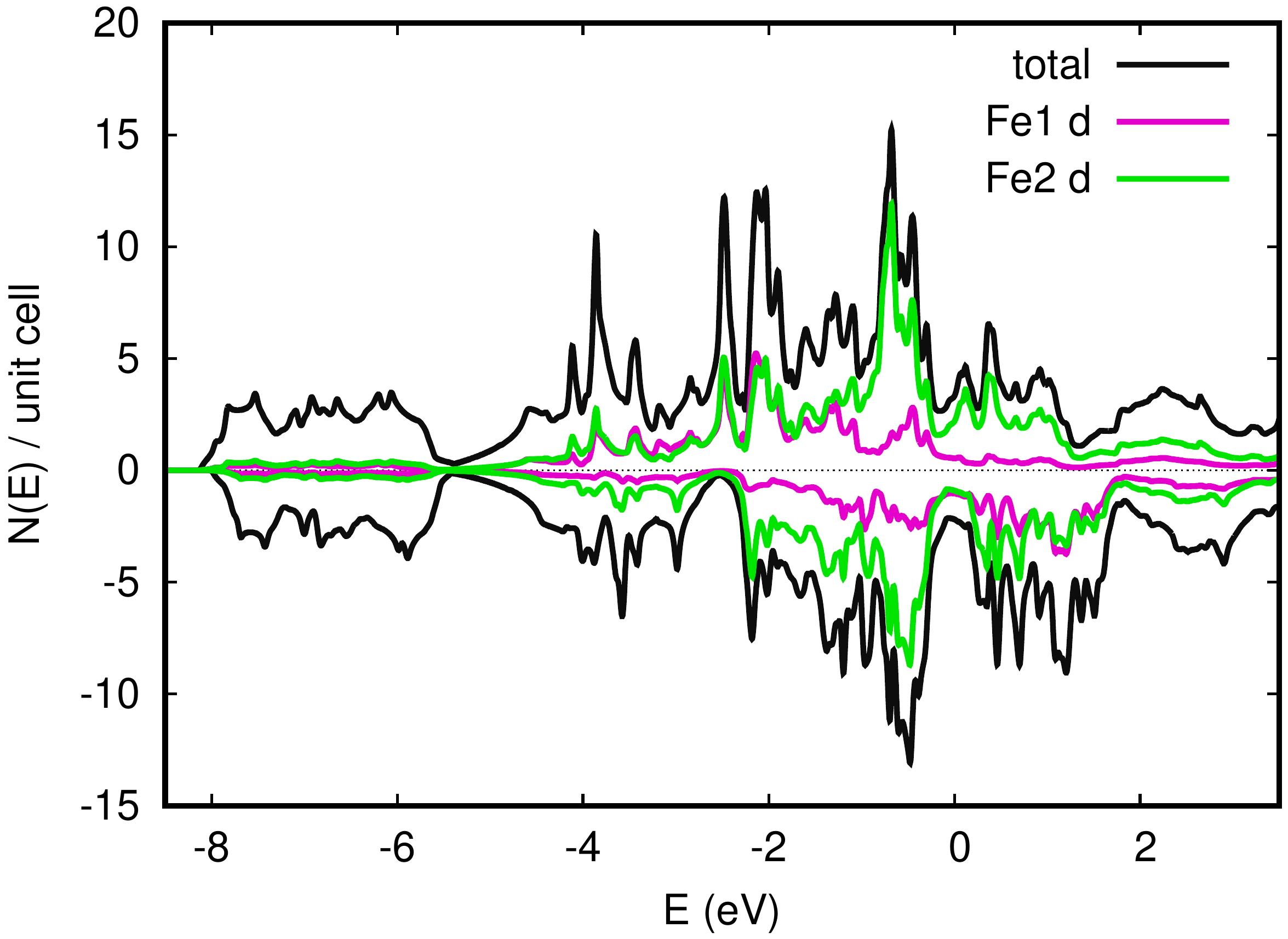}}
\caption{Total and partial electronic DOS plots shown on a per unit cell, two formula unit, basis.
Majority and minority spin DOS are shown above and below zero, respectively. The Fermi level is at zero energy.}
\label{fig:dos}
\end{figure}

We begin with the ground state properties. Lattice parameters and magnetic moments are summarized in
Table \ref{SCF_table}.
The differences between the LDA and PBE results are typical of DFT calculations.
The LDA gives somewhat lower magnetic moments and magnetization than the PBE, specifically
a magnetization of 0.71 $\mu_B$ vs. 0.78 $\mu_B$ per Fe at the
experimental lattice parameter. It also predicts smaller lattice parameters than PBE, with an equilibrium cell
volume that is 7\%
smaller.
It may also be noted that the calculated $c$/$a$ ratios (1.49 for PBE and 1.47 for LDA)
are slightly larger than the experimental value
of 1.46.
In any case, the magnitude and direction of the differences in lattice parameters between the LDA and PBE functionals
is typical of the behavior of these functionals in weakly and moderately correlated metals, including bulk Fe.
\cite{zhang,fu2}
As mentioned, the experimental magnetization is sample dependent. However, it is generally
somewhat smaller than our result. 
For example,
a recent experimental study reported 
0.57 $\mu_{B}$ per Fe for Ge$_{0.97}$Fe$_{3.03}$N$_{0.56}$.
\cite{Scholz2017}
This may be a consequence of differences in stoichiometry or DFT errors.
Interestingly, the type of deviation, where the experimental magnetization is smaller
than the calculated LDA magnetization, is characteristic of some itinerant magnetic
systems that are near quantum critical points.
In particular, standard density functional calculations, as with the LDA or PBE GGA,
do not include renormalization of the magnetization due to spin-fluctuations near the
critical point.
\cite{shimizu,aguayo,larson}
It may be of interest to experimentally study Fe$_3$GeN in this context as well as to perform detailed experimental
studies of the connection between magnetic behavior and nitrogen stoichiometry.

\begin{figure}[tbp]
\includegraphics[width=\columnwidth]{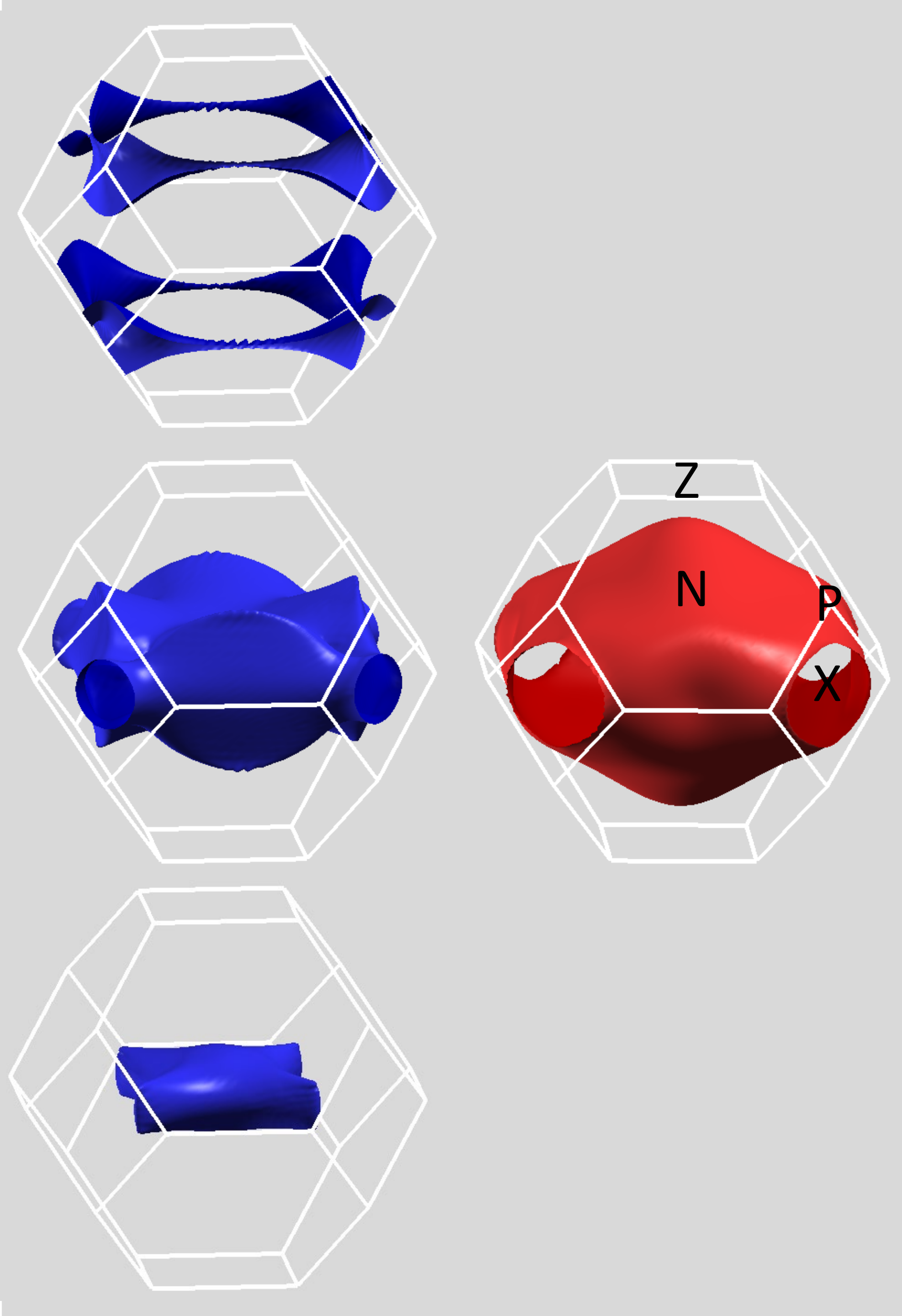}
\caption{Calculated Fermi surfaces for the PBE functional at the experimental
lattice parameters. There are three majority spin Fermi surfaces, shown in blue (left), and one
minority spin surface, shown in red (right).
$\Gamma$ is at the center of the body centered tetragonal zone as shown.
Other symmetry points corresponding to the points in the band
structure are as indicated on the minority spin surface (right).
With respect to the conventional tetragonal zone, Z is along the $k_z$ [0,0,1] direction,
while X is along the conventional cell [1,1,0] direction.}
\label{fig:Fermi}
\end{figure}

Turning to the details of the ground state
magnetic behavior, one notes that the two inequivalent Fe sites behave very
differently.
The magnetization is mainly from the Fe1, apical Fe, site,
while the in-plane Fe2 site has a much smaller moment.
As mentioned, the Fe2 site differs from the Fe1 site in that it has bent N-Fe-N bonds due to 
the rotation.
The magnetic moment of each Fe1 atom almost six times larger than that of an Fe2 atom.
In our calculations, while the Fe2 moment is small, it is aligned ferromagnetically with the Fe1 moment.

Fig. \ref{fig:band} shows the band structures for majority and minority spin.
These are for the PBE functional at the experimental lattice parameters.
The corresponding density of states (DOS) is given in Fig. \ref{fig:dos}.
Our calculated DOS is in accord with the prior calculation of Liu and
coworkers. \cite{Liu2020}
The Fermi level falls in a dip in the DOS. 
Fe $d$ states dominate the DOS between -4 eV and 2 eV, relative to the Fermi level.
There is considerable hybridization evident in the DOS as seen also in prior work, \cite{Liu2020}
and the projected band structure presented in the Appendix.
As mentioned, the magnetic behaviors of the Fe1 and Fe2 sites are different.
The analysis is somewhat complicated by the hybridization of the Fe1 and Fe2 states.
Nonetheless, the Fe1 and Fe2 projections of the DOS are significantly different from each other.
In particular, the Fe2 $d$ DOS, besides showing smaller exchange splitting, corresponding to the lower
moment, appears more narrow, in particular for the peak that occurs just below the Fermi level.
The density of states at the Fermi level, $N(E_F)$ are 3.32 eV$^{-1}$ and 2.19 eV$^{-1}$ per
six Fe atom unit cell, 
for majority and minority spin, respectively.
Fat band plots for the region close to the
Fermi energy, showing the Fe1 and Fe2 $d$ characters, are given in the Appendix.

Several bands cross the Fermi level, as seen in the band structure.
These lead to large Fermi surfaces in both the majority and minority spin.
It is noteworthy that these bands are more dispersive in the $k_z$ direction, i.e. $\Gamma$-Z
than in the in plane direction.
The Fermi surfaces are shown in
Fig. \ref{fig:Fermi}.
The majority spin shows
a hole-like rings around the zone edges and a small electron-like surface around
the zone center,
in addition to a large surface. The
occupancies of the corresponding three partially filled majority spin bands
are 0.966, 0.319, and 0.045.
The minority spin has one large sheet of Fermi surface, corresponding to a band with an
occupancy of 0.670.
The difference between the Fe1 and Fe2 sites is also reflected in the Fermi surface.
In particular, it 
is significantly anisotropic between the $k_z$ and the in-plane $k_x$ and $k_y$ directions,
perpendicular to $k_z$.
The large surfaces are flattened along $k_z$, particularly for the majority spin.
This reflects the higher dispersion of the bands in this direction, leading to the expectation
that the electrical conductivity should be higher in the $c$-axis direction, relative to in-plane.

\begin{figure}[hbtp]
\centerline{\includegraphics[width=\columnwidth]{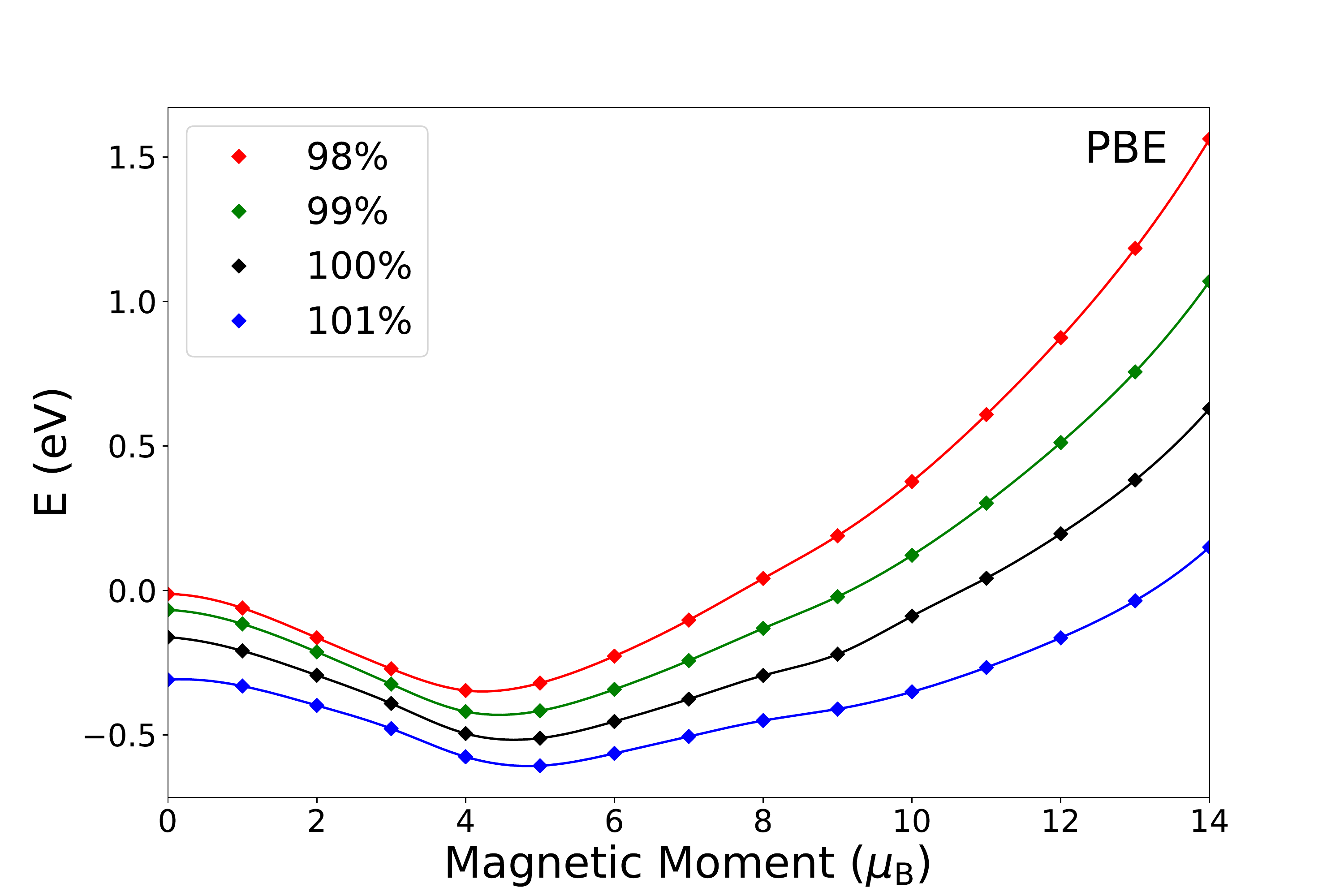}}
\centerline{\includegraphics[width=\columnwidth]{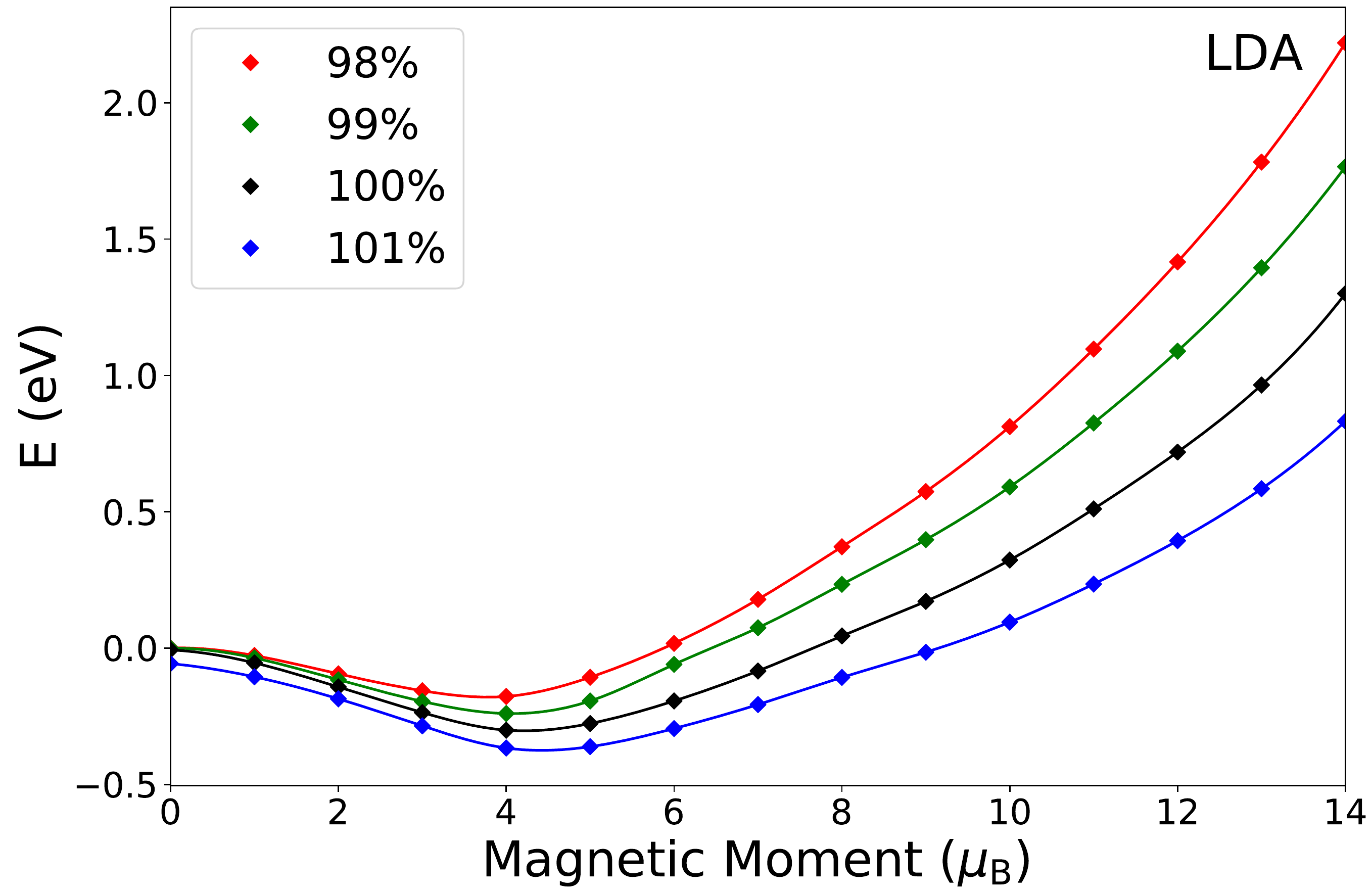}}
\caption{Calculated fixed spin moment magnetic energy as a function of magnetization on a per unit cell basis,
using the PBE and LDA functionals.
The energy zero is taken as the energy of a non-spin-polarized calculation for each set of lattice
parameters.
The constrained zero moment per unit cell calculation can have a lower
energy than the non-spin-polarized zero energy,
as a consequence of forming a lower energy state with canceling non-zero moments on the Fe1 and Fe2 sites.
This is seen at larger lattice parameters (99\% and larger for PBE and 101\% for LDA).
}
\label{fig:fsm}
\end{figure}

\begin{figure}[hbtp]
\centerline{\includegraphics[width=\columnwidth]{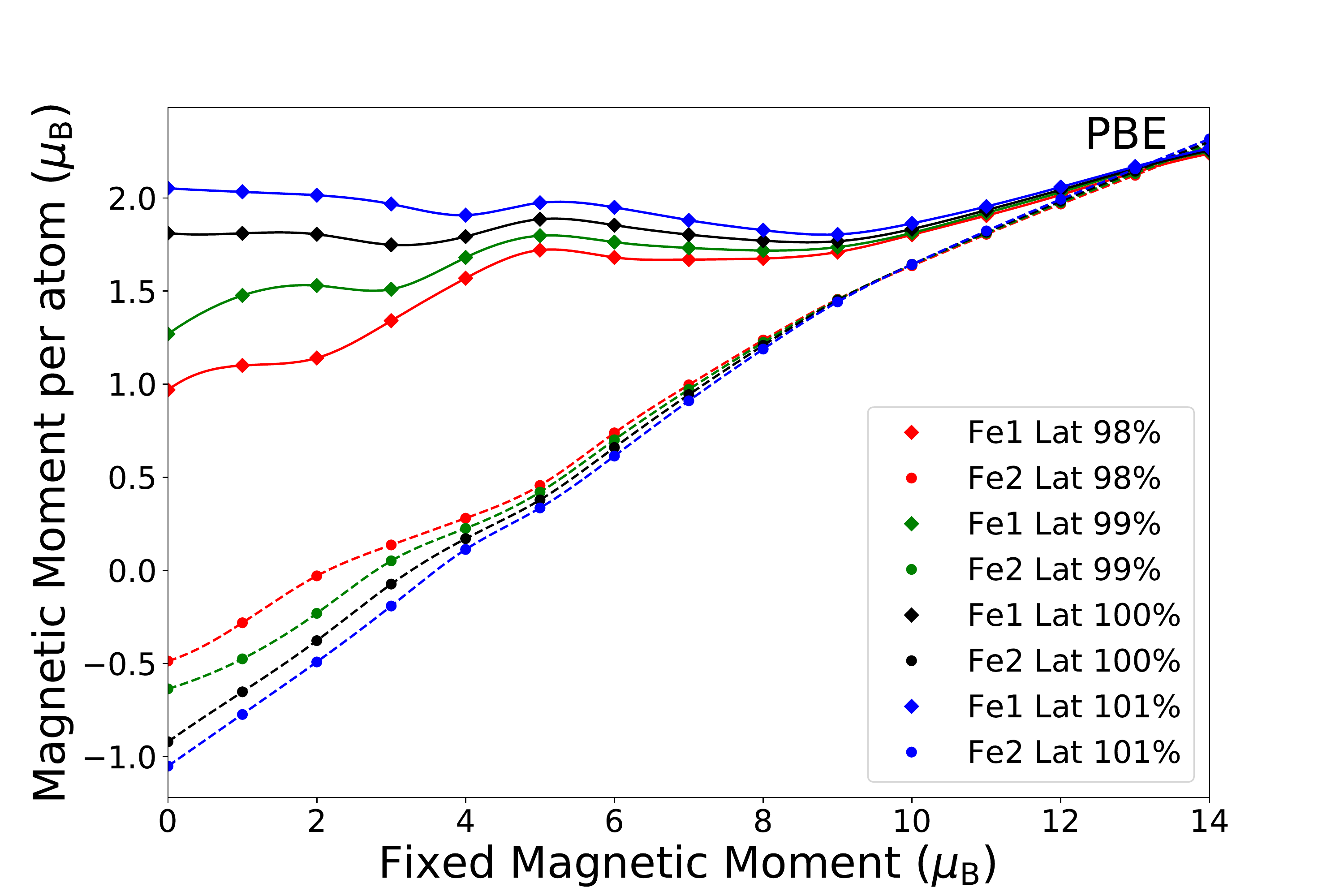}}
\centerline{\includegraphics[width=\columnwidth]{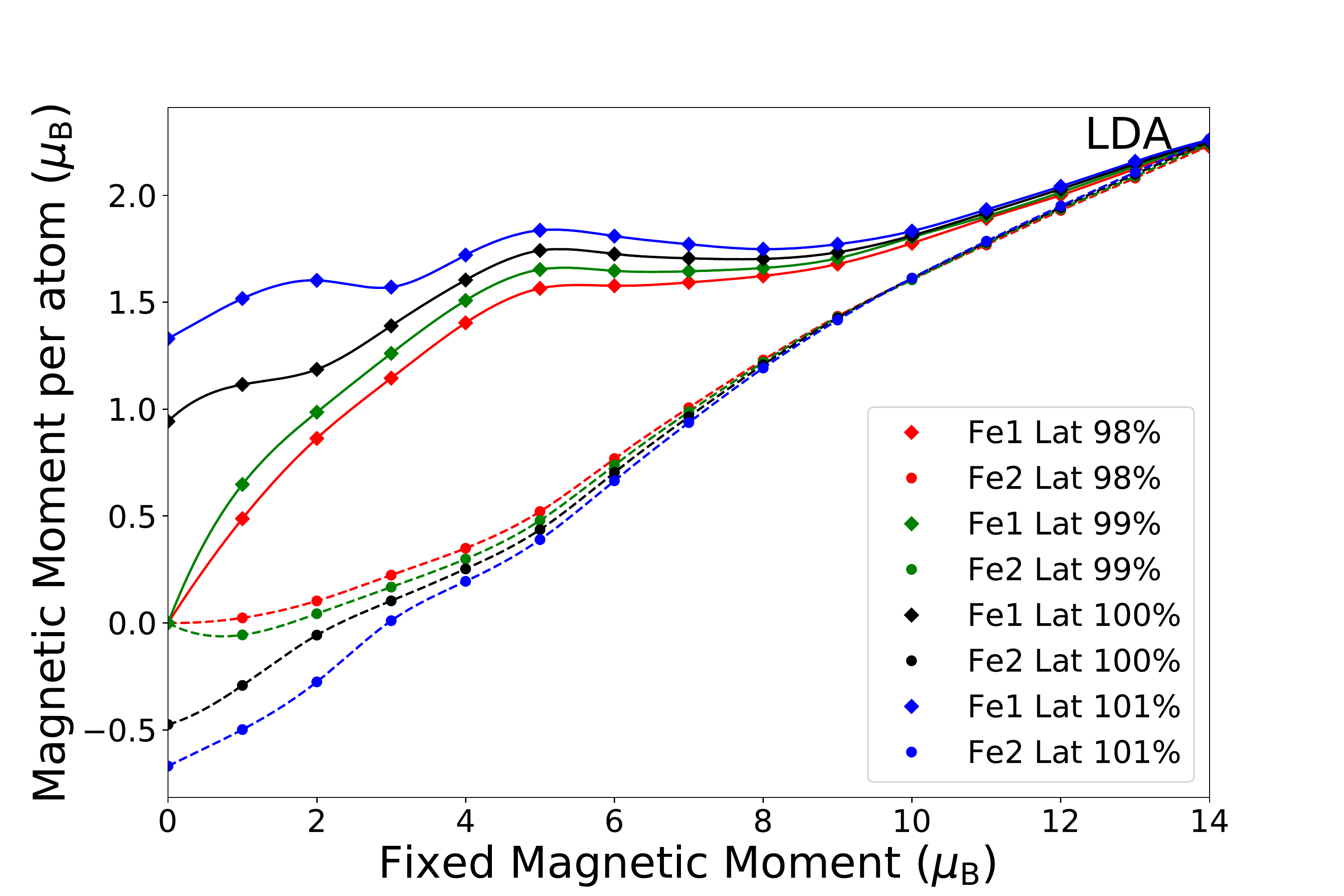}}
\caption{Fe1 and Fe2 moment on a per atom basis versus constrained magnetization per
unit cell from fixed spin moment calculations.}
\label{fig:Fe_fsm}
\end{figure}

We used constrained density functional theory, fixed spin moment
calculations to elucidate the nature of the magnetism.
\cite{fsm1,fsm2}
These were done both at the experimental lattice parameter, and under uniform compression and expansion,
for both the PBE and LDA functionals. We relaxed the internal coordinates of the Fe2 atom for each
magnetization and volume.
Fig. \ref{fig:fsm} gives the results of fixed spin moment calculations
for lattice parameters of 98\%, 99\%, 100\%, and 101\% of the experimental lattice values.
\citep{Kan2017}
The energies and magnetization are shown on a per unit cell basis.
The energy minimum for the experimental lattice
parameters (curves marked 100\% in comparison with the data
in the first set of columns in the table) corresponds to the self-consistent magnetization shown in
Table \ref{SCF_table}, as expected.
In addition, there is a flattening of the energy vs. magnetization in the vicinity of 10 $\mu_B$ per unit cell
($\sim$1.7 $mu_B$/Fe). This is with respect to the expected parabolic increase of the energy with magnetization
and
is particularly noticeable for the expanded cell and the PBE functional.
This suggests proximity to a metamagnetic transition from the ground state
to a high magnetization state, which might be
found in high magnetic field experiments.

Fig. \ref{fig:Fe_fsm} shows the magnetic moment of individual Fe1 and Fe2 atoms as a function of unit cell
magnetization
in the fixed spin moment calculations.
Although the Fe1 and Fe2 atoms are in chemically similar environments with similar coordination, they show
dramatically different behaviors.
The Fe2 moments are approximately linear with constrained magnetization for both functionals and
for the different lattice parameters. The Fe1 moments on the other hand, while they do vary, are much less
flexible.
This is in the sense that, particularly for the PBE functional and for the experimental and expanded lattice
parameter, the Fe1 moments do not show nearly as strong a dependence on magnetization as the Fe2 site.
Thus when the total magnetization is constrained to have a value different from the
equilibrium value the constraint is accommodated mainly by the Fe2 site.

Transition metal magnets are often characterized
according to the extent to which they show itinerant or local moment magnetism.
The local moment limit has stable moments, and the magnetic behavior is governed by the interactions between
the directions of these, as in the Heisenberg model, exemplified by many oxides, such as MnO.

The opposite itinerant limit is characterized by moments whose magnitude is variable and changes
for example with temperature, as for example in the Stoner picture, exemplified by fcc Ni.
Based on the results in Fig. \ref{fig:Fe_fsm}, the magnetic moment of the Fe2 goes up strongly
as the fixed spin moment increases in contrast to the behavior of the Fe1, which
is relatively invariant.
This means that the Fe1 atoms behave as if they have local moments,
while the Fe2 atoms show a more itinerant nature.
Thus, the magnetism of
Fe$_{3}$GeN could be described as local Fe1 moments interacting through an itinerant system comprised by the Fe2
atoms.
This result is robust. We find this differentiation of the Fe1 and Fe2 sites,
both with the LDA and PBE functionals, at the experimental and the
calculated lattice parameters and over a range of volumes around the experimental volume.

This nature could in principle be probed by neutron diffraction measurements of the Fe1 and Fe2 moments
as a function of temperature.
While Fe$_3$GeN does not have a particularly high Curie temperature, it is notable that
some very high Curie temperature materials, particularly
Heusler Co$_{2}$FeSi (T$_{c}$ $\simeq$ 1100 K) 
show a similar characteristic.
\cite{Qin2020}
In any case, the results show that in spite of their very similar chemical and coordination environments
the magnetic behaviors of the two Fe sites in Fe$_3$GeN are very different.
This reflects a sensitive interplay of structure and magnetism in an Fe compound.

\section{Summary and Conclusions}

We report a DFT investigation of the ternary iron nitride compound, Fe$_{3}$GeN, which has a distorted antiperovskite
structure, in which octahedral rotation leads to two distinct Fe sites.
Remarkably, this also leads to very different magnetic properties of these Fe atoms.
One site, the Fe1, apical site, accounts for most of the magnetization and has a relatively stable moment.
The other site, Fe2, which is the in-plane Fe site, has a moment that is aligned with the
Fe1 moment but is much smaller and much more flexible than the Fe1 moment.
The magnetism may then be described as local Fe1 moments embedded in an itinerant Fe2 background.
This separation into two sites with very different behavior is an unexpected deviation from the
rule of parsimony, according to which the different Fe atoms would be expected to behave similarly.

The results suggest that, analogously to oxide perovskite materials, modulation of magnetic properties via
chemical or strain control of octahedral rotation may be feasible.
This may then lead to approaches for tuning magnetism to realize properties of interest, for
example tuning magnetic transitions to quantum critical regimes or to proximity to metamagnetic
transitions of interest for devices.
It will be of strong interest to extract rational design principles, analogous to those in oxides, for
understanding and predicting the variation of structure with chemistry, order, strain and other
parameters for these intermetallic antiperovskite phases.

\section*{Appendix}

\begin{figure}[tb]
\includegraphics[width=\columnwidth]{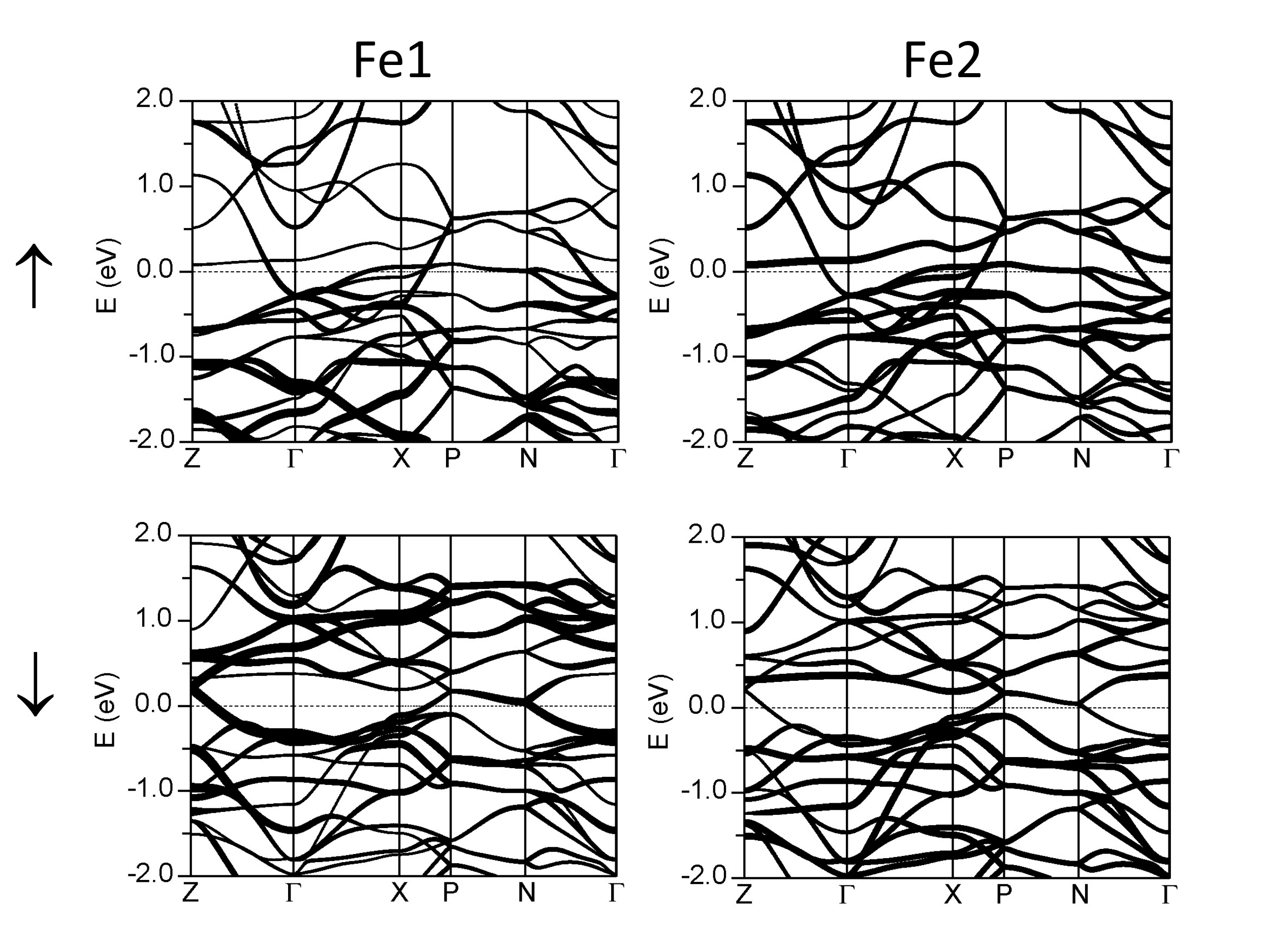}
\caption{Calculated projected band structures for majority (top) and minority spin (bottom).
The dashed line denotes the Fermi level at 0 eV. The projections are on a per atom basis.}
\label{fatbands}
\end{figure}

The projected band structures of Fe1 $d$ and Fe2 $d$ character
for majority and minority spin near the Fermi energy, $E_F$, are shown in fat-band representations in Fig.
\ref{fatbands}. The projections were onto the corresponding LAPW spheres.
As seen, the bands closest to $E_F$ have hybridized Fe d character from the two sites.

Fe2 makes larger contributions for the majority spin, while Fe1 makes larger contributions for the
minority spin in the energy range shown.
This may also be noted from the density of states.

\begin{acknowledgments}

We are grateful for helpful discussions with Bing-Hua Lei.
This work was supported by the Department of Energy, Office of Basic Energy Science, Award DE-SC0019114.

\end{acknowledgments}

\bibliography{Fe3GeN_ref}

\begin{thebibliography}{36}%
\makeatletter
\providecommand \@ifxundefined [1]{%
 \@ifx{#1\undefined}
}%
\providecommand \@ifnum [1]{%
 \ifnum #1\expandafter \@firstoftwo
 \else \expandafter \@secondoftwo
 \fi
}%
\providecommand \@ifx [1]{%
 \ifx #1\expandafter \@firstoftwo
 \else \expandafter \@secondoftwo
 \fi
}%
\providecommand \natexlab [1]{#1}%
\providecommand \enquote  [1]{``#1''}%
\providecommand \bibnamefont  [1]{#1}%
\providecommand \bibfnamefont [1]{#1}%
\providecommand \citenamefont [1]{#1}%
\providecommand \href@noop [0]{\@secondoftwo}%
\providecommand \href [0]{\begingroup \@sanitize@url \@href}%
\providecommand \@href[1]{\@@startlink{#1}\@@href}%
\providecommand \@@href[1]{\endgroup#1\@@endlink}%
\providecommand \@sanitize@url [0]{\catcode `\\12\catcode `\$12\catcode
  `\&12\catcode `\#12\catcode `\^12\catcode `\_12\catcode `\%12\relax}%
\providecommand \@@startlink[1]{}%
\providecommand \@@endlink[0]{}%
\providecommand \url  [0]{\begingroup\@sanitize@url \@url }%
\providecommand \@url [1]{\endgroup\@href {#1}{\urlprefix }}%
\providecommand \urlprefix  [0]{URL }%
\providecommand \Eprint [0]{\href }%
\providecommand \doibase [0]{http://dx.doi.org/}%
\providecommand \selectlanguage [0]{\@gobble}%
\providecommand \bibinfo  [0]{\@secondoftwo}%
\providecommand \bibfield  [0]{\@secondoftwo}%
\providecommand \translation [1]{[#1]}%
\providecommand \BibitemOpen [0]{}%
\providecommand \bibitemStop [0]{}%
\providecommand \bibitemNoStop [0]{.\EOS\space}%
\providecommand \EOS [0]{\spacefactor3000\relax}%
\providecommand \BibitemShut  [1]{\csname bibitem#1\endcsname}%
\let\auto@bib@innerbib\@empty
\bibitem [{\citenamefont {Jack}(1948)}]{Jack1948}%
  \BibitemOpen
  \bibfield  {author} {\bibinfo {author} {\bibfnamefont {K.~H.}\ \bibnamefont
  {Jack}},\ }\href {\doibase 10.1098/rspa.1948.0100} {\bibfield  {journal}
  {\bibinfo  {journal} {Proc. R. Soc. Lond}\ }\textbf {\bibinfo {volume}
  {195}},\ \bibinfo {pages} {34} (\bibinfo {year} {1948})}\BibitemShut
  {NoStop}%
\bibitem [{\citenamefont {Takahashi}\ \emph {et~al.}(1994)\citenamefont
  {Takahashi}, \citenamefont {Shoji}, \citenamefont {Takahashi}, \citenamefont
  {Nashi},\ and\ \citenamefont {Wakiyama}}]{takahashi}%
  \BibitemOpen
  \bibfield  {author} {\bibinfo {author} {\bibfnamefont {M.}~\bibnamefont
  {Takahashi}}, \bibinfo {author} {\bibfnamefont {H.}~\bibnamefont {Shoji}},
  \bibinfo {author} {\bibfnamefont {H.}~\bibnamefont {Takahashi}}, \bibinfo
  {author} {\bibfnamefont {H.}~\bibnamefont {Nashi}}, \ and\ \bibinfo {author}
  {\bibfnamefont {T.}~\bibnamefont {Wakiyama}},\ }\href@noop {} {\bibfield
  {journal} {\bibinfo  {journal} {J. Appl. Phys.}\ }\textbf {\bibinfo {volume}
  {76}},\ \bibinfo {pages} {6642} (\bibinfo {year} {1994})}\BibitemShut
  {NoStop}%
\bibitem [{\citenamefont {Wang}(2020)}]{Wang2020}%
  \BibitemOpen
  \bibfield  {author} {\bibinfo {author} {\bibfnamefont {J.-P.}\ \bibnamefont
  {Wang}},\ }\href {\doibase https://doi.org/10.1016/j.jmmm.2019.165962}
  {\bibfield  {journal} {\bibinfo  {journal} {Journal of Magnetism and Magnetic
  Materials}\ }\textbf {\bibinfo {volume} {497}},\ \bibinfo {pages} {165962}
  (\bibinfo {year} {2020})}\BibitemShut {NoStop}%
\bibitem [{\citenamefont {Bramfitt}(1998)}]{bramfitt}%
  \BibitemOpen
  \bibfield  {author} {\bibinfo {author} {\bibfnamefont {B.~L.}\ \bibnamefont
  {Bramfitt}},\ }in\ \href@noop {} {\emph {\bibinfo {booktitle} {Metals
  Handbook, Desk Edition, 2nd. Ed.}}},\ \bibinfo {editor} {edited by\ \bibinfo
  {editor} {\bibfnamefont {J.~R.}\ \bibnamefont {Davis}}}\ (\bibinfo
  {publisher} {ASM International},\ \bibinfo {address} {Materials Park},\
  \bibinfo {year} {1998})\ pp.\ \bibinfo {pages} {153--173}\BibitemShut
  {NoStop}%
\bibitem [{\citenamefont {Fu}\ and\ \citenamefont {Singh}(2018)}]{fu}%
  \BibitemOpen
  \bibfield  {author} {\bibinfo {author} {\bibfnamefont {Y.}~\bibnamefont
  {Fu}}\ and\ \bibinfo {author} {\bibfnamefont {D.~J.}\ \bibnamefont {Singh}},\
  }\href@noop {} {\bibfield  {journal} {\bibinfo  {journal} {Phys. Rev. Lett.}\
  }\textbf {\bibinfo {volume} {121}},\ \bibinfo {pages} {207201} (\bibinfo
  {year} {2018})}\BibitemShut {NoStop}%
\bibitem [{\citenamefont {Zieschang}\ \emph {et~al.}(2017)\citenamefont
  {Zieschang}, \citenamefont {Bocarsly}, \citenamefont {Durrshnabel},
  \citenamefont {{Monica-Luna}}, \citenamefont {Kleebe}, \citenamefont
  {Seshadri},\ and\ \citenamefont {Albert}}]{zieschang}%
  \BibitemOpen
  \bibfield  {author} {\bibinfo {author} {\bibfnamefont {A.~M.}\ \bibnamefont
  {Zieschang}}, \bibinfo {author} {\bibfnamefont {J.~D.}\ \bibnamefont
  {Bocarsly}}, \bibinfo {author} {\bibfnamefont {M.}~\bibnamefont
  {Durrshnabel}}, \bibinfo {author} {\bibfnamefont {L.}~\bibnamefont
  {{Monica-Luna}}}, \bibinfo {author} {\bibfnamefont {H.~J.}\ \bibnamefont
  {Kleebe}}, \bibinfo {author} {\bibfnamefont {R.}~\bibnamefont {Seshadri}}, \
  and\ \bibinfo {author} {\bibfnamefont {B.}~\bibnamefont {Albert}},\
  }\href@noop {} {\bibfield  {journal} {\bibinfo  {journal} {Chem. Mater.}\
  }\textbf {\bibinfo {volume} {29}},\ \bibinfo {pages} {621} (\bibinfo {year}
  {2017})}\BibitemShut {NoStop}%
\bibitem [{\citenamefont {Chen}\ \emph {et~al.}(1991)\citenamefont {Chen},
  \citenamefont {Jin}, \citenamefont {Tiefel}, \citenamefont {Hsieh},
  \citenamefont {Gyorgy},\ and\ \citenamefont {{Johnson, Jr.}}}]{chen}%
  \BibitemOpen
  \bibfield  {author} {\bibinfo {author} {\bibfnamefont {S.~K.}\ \bibnamefont
  {Chen}}, \bibinfo {author} {\bibfnamefont {S.}~\bibnamefont {Jin}}, \bibinfo
  {author} {\bibfnamefont {T.~H.}\ \bibnamefont {Tiefel}}, \bibinfo {author}
  {\bibfnamefont {Y.~F.}\ \bibnamefont {Hsieh}}, \bibinfo {author}
  {\bibfnamefont {E.~M.}\ \bibnamefont {Gyorgy}}, \ and\ \bibinfo {author}
  {\bibfnamefont {D.~W.}\ \bibnamefont {{Johnson, Jr.}}},\ }\href@noop {}
  {\bibfield  {journal} {\bibinfo  {journal} {J. Appl. Phys.}\ }\textbf
  {\bibinfo {volume} {70}},\ \bibinfo {pages} {6247} (\bibinfo {year}
  {1991})}\BibitemShut {NoStop}%
\bibitem [{\citenamefont {{Costa-Kramer}}\ \emph {et~al.}(2004)\citenamefont
  {{Costa-Kramer}}, \citenamefont {Borsa}, \citenamefont {{Garcia-Martin}},
  \citenamefont {{Martin-Gonzalez}}, \citenamefont {Boerma},\ and\
  \citenamefont {Briones}}]{costa-kramer}%
  \BibitemOpen
  \bibfield  {author} {\bibinfo {author} {\bibfnamefont {J.~L.}\ \bibnamefont
  {{Costa-Kramer}}}, \bibinfo {author} {\bibfnamefont {D.~M.}\ \bibnamefont
  {Borsa}}, \bibinfo {author} {\bibfnamefont {J.~M.}\ \bibnamefont
  {{Garcia-Martin}}}, \bibinfo {author} {\bibfnamefont {M.~S.}\ \bibnamefont
  {{Martin-Gonzalez}}}, \bibinfo {author} {\bibfnamefont {D.~O.}\ \bibnamefont
  {Boerma}}, \ and\ \bibinfo {author} {\bibfnamefont {F.}~\bibnamefont
  {Briones}},\ }\href@noop {} {\bibfield  {journal} {\bibinfo  {journal} {Phys.
  Rev. B}\ }\textbf {\bibinfo {volume} {69}},\ \bibinfo {pages} {144402}
  (\bibinfo {year} {2004})}\BibitemShut {NoStop}%
\bibitem [{\citenamefont {Burghaus}\ \emph {et~al.}(2010)\citenamefont
  {Burghaus}, \citenamefont {Wessel}, \citenamefont {Houben},\ and\
  \citenamefont {Dronskowski}}]{burghaus}%
  \BibitemOpen
  \bibfield  {author} {\bibinfo {author} {\bibfnamefont {J.}~\bibnamefont
  {Burghaus}}, \bibinfo {author} {\bibfnamefont {M.}~\bibnamefont {Wessel}},
  \bibinfo {author} {\bibfnamefont {A.}~\bibnamefont {Houben}}, \ and\ \bibinfo
  {author} {\bibfnamefont {R.}~\bibnamefont {Dronskowski}},\ }\href@noop {}
  {\bibfield  {journal} {\bibinfo  {journal} {Inorg. Chem.}\ }\textbf {\bibinfo
  {volume} {49}},\ \bibinfo {pages} {10148} (\bibinfo {year}
  {2010})}\BibitemShut {NoStop}%
\bibitem [{\citenamefont {Houben}\ \emph {et~al.}(2005)\citenamefont {Houben},
  \citenamefont {Muller}, \citenamefont {Appen}, \citenamefont {Lueken},
  \citenamefont {Dr},\ and\ \citenamefont {Dronskowski}}]{Houben2006}%
  \BibitemOpen
  \bibfield  {author} {\bibinfo {author} {\bibfnamefont {A.}~\bibnamefont
  {Houben}}, \bibinfo {author} {\bibfnamefont {P.}~\bibnamefont {Muller}},
  \bibinfo {author} {\bibfnamefont {J.}~\bibnamefont {Appen}}, \bibinfo
  {author} {\bibfnamefont {H.}~\bibnamefont {Lueken}}, \bibinfo {author}
  {\bibfnamefont {R.}~\bibnamefont {Dr}}, \ and\ \bibinfo {author}
  {\bibfnamefont {R.}~\bibnamefont {Dronskowski}},\ }\href@noop {} {\bibfield
  {journal} {\bibinfo  {journal} {Angew. Chem.}\ }\textbf {\bibinfo {volume}
  {44}},\ \bibinfo {pages} {7212} (\bibinfo {year} {2005})}\BibitemShut
  {NoStop}%
\bibitem [{\citenamefont {Fu}\ \emph {et~al.}(2015)\citenamefont {Fu},
  \citenamefont {Lin},\ and\ \citenamefont {Wang}}]{fu-2014}%
  \BibitemOpen
  \bibfield  {author} {\bibinfo {author} {\bibfnamefont {Y.}~\bibnamefont
  {Fu}}, \bibinfo {author} {\bibfnamefont {S.}~\bibnamefont {Lin}}, \ and\
  \bibinfo {author} {\bibfnamefont {B.}~\bibnamefont {Wang}},\ }\href@noop {}
  {\bibfield  {journal} {\bibinfo  {journal} {J. Magn. Magn. Mater.}\ }\textbf
  {\bibinfo {volume} {378}},\ \bibinfo {pages} {54} (\bibinfo {year}
  {2015})}\BibitemShut {NoStop}%
\bibitem [{\citenamefont {Niewa}(2019)}]{niewa}%
  \BibitemOpen
  \bibfield  {author} {\bibinfo {author} {\bibfnamefont {R.}~\bibnamefont
  {Niewa}},\ }\href@noop {} {\bibfield  {journal} {\bibinfo  {journal} {Eur. J.
  Inorg. Chem.}\ }\textbf {\bibinfo {volume} {2019}},\ \bibinfo {pages} {3647}
  (\bibinfo {year} {2019})}\BibitemShut {NoStop}%
\bibitem [{\citenamefont {Glazer}(1972)}]{glazer}%
  \BibitemOpen
  \bibfield  {author} {\bibinfo {author} {\bibfnamefont {A.~M.}\ \bibnamefont
  {Glazer}},\ }\href@noop {} {\bibfield  {journal} {\bibinfo  {journal} {Acta
  Cryst.}\ }\textbf {\bibinfo {volume} {B28}},\ \bibinfo {pages} {3384}
  (\bibinfo {year} {1972})}\BibitemShut {NoStop}%
\bibitem [{\citenamefont {He}\ \emph {et~al.}(2010)\citenamefont {He},
  \citenamefont {Borisevich}, \citenamefont {Kalinin}, \citenamefont
  {Pennycook},\ and\ \citenamefont {Pantelides}}]{he}%
  \BibitemOpen
  \bibfield  {author} {\bibinfo {author} {\bibfnamefont {J.}~\bibnamefont
  {He}}, \bibinfo {author} {\bibfnamefont {A.}~\bibnamefont {Borisevich}},
  \bibinfo {author} {\bibfnamefont {S.~V.}\ \bibnamefont {Kalinin}}, \bibinfo
  {author} {\bibfnamefont {S.~J.}\ \bibnamefont {Pennycook}}, \ and\ \bibinfo
  {author} {\bibfnamefont {S.~T.}\ \bibnamefont {Pantelides}},\ }\href@noop {}
  {\bibfield  {journal} {\bibinfo  {journal} {Phys. Rev. Lett.}\ }\textbf
  {\bibinfo {volume} {105}},\ \bibinfo {pages} {227203} (\bibinfo {year}
  {2010})}\BibitemShut {NoStop}%
\bibitem [{\citenamefont {Rondinelli}\ and\ \citenamefont
  {Fennie}(2012)}]{rondinelli}%
  \BibitemOpen
  \bibfield  {author} {\bibinfo {author} {\bibfnamefont {J.~M.}\ \bibnamefont
  {Rondinelli}}\ and\ \bibinfo {author} {\bibfnamefont {C.~J.}\ \bibnamefont
  {Fennie}},\ }\href@noop {} {\bibfield  {journal} {\bibinfo  {journal} {Adv.
  Mater.}\ }\textbf {\bibinfo {volume} {24}},\ \bibinfo {pages} {1961}
  (\bibinfo {year} {2012})}\BibitemShut {NoStop}%
\bibitem [{\citenamefont {Bilc}\ and\ \citenamefont {Singh}(2006)}]{bilc}%
  \BibitemOpen
  \bibfield  {author} {\bibinfo {author} {\bibfnamefont {D.~I.}\ \bibnamefont
  {Bilc}}\ and\ \bibinfo {author} {\bibfnamefont {D.~J.}\ \bibnamefont
  {Singh}},\ }\href@noop {} {\bibfield  {journal} {\bibinfo  {journal} {Phys.
  Rev. Lett.}\ }\textbf {\bibinfo {volume} {96}},\ \bibinfo {pages} {147602}
  (\bibinfo {year} {2006})}\BibitemShut {NoStop}%
\bibitem [{\citenamefont {Scholz}\ and\ \citenamefont
  {Dronskowski}(2017)}]{Scholz2017}%
  \BibitemOpen
  \bibfield  {author} {\bibinfo {author} {\bibfnamefont {T.}~\bibnamefont
  {Scholz}}\ and\ \bibinfo {author} {\bibfnamefont {R.}~\bibnamefont
  {Dronskowski}},\ }\href {\doibase 10.1039/C6TC04543J} {\bibfield  {journal}
  {\bibinfo  {journal} {J. Mater. Chem. C}\ }\textbf {\bibinfo {volume} {5}},\
  \bibinfo {pages} {166} (\bibinfo {year} {2017})}\BibitemShut {NoStop}%
\bibitem [{\citenamefont {Liu}\ \emph {et~al.}(2020)\citenamefont {Liu},
  \citenamefont {Zhang}, \citenamefont {Kan}, \citenamefont {Liu},
  \citenamefont {Feng}, \citenamefont {Yang}, \citenamefont {Lv}, \citenamefont
  {Hu},\ and\ \citenamefont {Shezad}}]{Liu2020}%
  \BibitemOpen
  \bibfield  {author} {\bibinfo {author} {\bibfnamefont {C.}~\bibnamefont
  {Liu}}, \bibinfo {author} {\bibfnamefont {C.}~\bibnamefont {Zhang}}, \bibinfo
  {author} {\bibfnamefont {X.}~\bibnamefont {Kan}}, \bibinfo {author}
  {\bibfnamefont {X.}~\bibnamefont {Liu}}, \bibinfo {author} {\bibfnamefont
  {S.}~\bibnamefont {Feng}}, \bibinfo {author} {\bibfnamefont {Y.}~\bibnamefont
  {Yang}}, \bibinfo {author} {\bibfnamefont {Q.}~\bibnamefont {Lv}}, \bibinfo
  {author} {\bibfnamefont {J.}~\bibnamefont {Hu}}, \ and\ \bibinfo {author}
  {\bibfnamefont {M.}~\bibnamefont {Shezad}},\ }\href
  {https://doi.org/10.1021/acs.jpcc.0c00307} {\bibfield  {journal} {\bibinfo
  {journal} {J. Phys. Chem. C}\ }\textbf {\bibinfo {volume} {124}},\ \bibinfo
  {pages} {6321} (\bibinfo {year} {2020})}\BibitemShut {NoStop}%
\bibitem [{\citenamefont {Boller}(1968)}]{boller1968}%
  \BibitemOpen
  \bibfield  {author} {\bibinfo {author} {\bibfnamefont {H.}~\bibnamefont
  {Boller}},\ }\href@noop {} {\bibfield  {journal} {\bibinfo  {journal}
  {Monatshefte f{\"u}r Chemie}\ }\textbf {\bibinfo {volume} {99}},\ \bibinfo
  {pages} {2444} (\bibinfo {year} {1968})}\BibitemShut {NoStop}%
\bibitem [{\citenamefont {Kan}\ \emph {et~al.}(2017)\citenamefont {Kan},
  \citenamefont {Wang}, \citenamefont {Zhang}, \citenamefont {Zu},
  \citenamefont {Lin}, \citenamefont {Lin}, \citenamefont {Tong}, \citenamefont
  {Song},\ and\ \citenamefont {Sun}}]{Kan2017}%
  \BibitemOpen
  \bibfield  {author} {\bibinfo {author} {\bibfnamefont {X.~C.}\ \bibnamefont
  {Kan}}, \bibinfo {author} {\bibfnamefont {B.~S.}\ \bibnamefont {Wang}},
  \bibinfo {author} {\bibfnamefont {L.}~\bibnamefont {Zhang}}, \bibinfo
  {author} {\bibfnamefont {L.}~\bibnamefont {Zu}}, \bibinfo {author}
  {\bibfnamefont {S.}~\bibnamefont {Lin}}, \bibinfo {author} {\bibfnamefont
  {J.~C.}\ \bibnamefont {Lin}}, \bibinfo {author} {\bibfnamefont
  {P.}~\bibnamefont {Tong}}, \bibinfo {author} {\bibfnamefont {W.~H.}\
  \bibnamefont {Song}}, \ and\ \bibinfo {author} {\bibfnamefont {Y.~P.}\
  \bibnamefont {Sun}},\ }\href {\doibase 10.1039/C6CP08020K} {\bibfield
  {journal} {\bibinfo  {journal} {Phys. Chem. Chem. Phys.}\ }\textbf {\bibinfo
  {volume} {19}},\ \bibinfo {pages} {13703} (\bibinfo {year}
  {2017})}\BibitemShut {NoStop}%
\bibitem [{\citenamefont {Kan}\ \emph {et~al.}(2016)\citenamefont {Kan},
  \citenamefont {Wang}, \citenamefont {Zu}, \citenamefont {Lin}, \citenamefont
  {Lin}, \citenamefont {Tong}, \citenamefont {Song},\ and\ \citenamefont
  {Sun}}]{Kan2016}%
  \BibitemOpen
  \bibfield  {author} {\bibinfo {author} {\bibfnamefont {X.~C.}\ \bibnamefont
  {Kan}}, \bibinfo {author} {\bibfnamefont {B.~S.}\ \bibnamefont {Wang}},
  \bibinfo {author} {\bibfnamefont {L.}~\bibnamefont {Zu}}, \bibinfo {author}
  {\bibfnamefont {S.}~\bibnamefont {Lin}}, \bibinfo {author} {\bibfnamefont
  {J.~C.}\ \bibnamefont {Lin}}, \bibinfo {author} {\bibfnamefont
  {P.}~\bibnamefont {Tong}}, \bibinfo {author} {\bibfnamefont {W.~H.}\
  \bibnamefont {Song}}, \ and\ \bibinfo {author} {\bibfnamefont {Y.~P.}\
  \bibnamefont {Sun}},\ }\href {\doibase 10.1039/C6RA15976A} {\bibfield
  {journal} {\bibinfo  {journal} {RSC Adv.}\ }\textbf {\bibinfo {volume} {6}},\
  \bibinfo {pages} {104433} (\bibinfo {year} {2016})}\BibitemShut {NoStop}%
\bibitem [{\citenamefont {Mohn}\ \emph {et~al.}(1992)\citenamefont {Mohn},
  \citenamefont {Schwarz}, \citenamefont {Matar},\ and\ \citenamefont
  {Demazeau}}]{Mohn1992}%
  \BibitemOpen
  \bibfield  {author} {\bibinfo {author} {\bibfnamefont {P.}~\bibnamefont
  {Mohn}}, \bibinfo {author} {\bibfnamefont {K.}~\bibnamefont {Schwarz}},
  \bibinfo {author} {\bibfnamefont {S.}~\bibnamefont {Matar}}, \ and\ \bibinfo
  {author} {\bibfnamefont {G.}~\bibnamefont {Demazeau}},\ }\href {\doibase
  10.1103/PhysRevB.45.4000} {\bibfield  {journal} {\bibinfo  {journal} {Phys.
  Rev. B}\ }\textbf {\bibinfo {volume} {45}},\ \bibinfo {pages} {4000}
  (\bibinfo {year} {1992})}\BibitemShut {NoStop}%
\bibitem [{\citenamefont {Rebaza}\ \emph {et~al.}(2011)\citenamefont {Rebaza},
  \citenamefont {Desimoni}, \citenamefont {Kurian}, \citenamefont
  {Bhattacharyya}, \citenamefont {Gajbhiye},\ and\ \citenamefont
  {Blanc{\'a}}}]{Rebaza2011}%
  \BibitemOpen
  \bibfield  {author} {\bibinfo {author} {\bibfnamefont {A.~V.~G.}\
  \bibnamefont {Rebaza}}, \bibinfo {author} {\bibfnamefont {J.}~\bibnamefont
  {Desimoni}}, \bibinfo {author} {\bibfnamefont {S.}~\bibnamefont {Kurian}},
  \bibinfo {author} {\bibfnamefont {S.}~\bibnamefont {Bhattacharyya}}, \bibinfo
  {author} {\bibfnamefont {N.~S.}\ \bibnamefont {Gajbhiye}}, \ and\ \bibinfo
  {author} {\bibfnamefont {E.~P.~Y.}\ \bibnamefont {Blanc{\'a}}},\ }\href@noop
  {} {\bibfield  {journal} {\bibinfo  {journal} {J. Phys. Chem. C}\ }\textbf
  {\bibinfo {volume} {115}},\ \bibinfo {pages} {23081} (\bibinfo {year}
  {2011})}\BibitemShut {NoStop}%
\bibitem [{\citenamefont {Singh}\ and\ \citenamefont
  {Nordstrom}(2006)}]{wien2klapw}%
  \BibitemOpen
  \bibfield  {author} {\bibinfo {author} {\bibfnamefont {D.~J.}\ \bibnamefont
  {Singh}}\ and\ \bibinfo {author} {\bibfnamefont {L.}~\bibnamefont
  {Nordstrom}},\ }\href@noop {} {\emph {\bibinfo {title} {Planewaves,
  Pseudopotentials, and the LAPW method, 2nd Ed.}}}\ (\bibinfo  {publisher}
  {Springer, Berlin},\ \bibinfo {year} {2006})\BibitemShut {NoStop}%
\bibitem [{\citenamefont {Blaha}\ \emph {et~al.}(2001)\citenamefont {Blaha},
  \citenamefont {Schwarz}, \citenamefont {Madsen}, \citenamefont {Kvasnicka},\
  and\ \citenamefont {Luitz}}]{wien2kcode}%
  \BibitemOpen
  \bibfield  {author} {\bibinfo {author} {\bibfnamefont {P.}~\bibnamefont
  {Blaha}}, \bibinfo {author} {\bibfnamefont {K.}~\bibnamefont {Schwarz}},
  \bibinfo {author} {\bibfnamefont {G.}~\bibnamefont {Madsen}}, \bibinfo
  {author} {\bibfnamefont {D.}~\bibnamefont {Kvasnicka}}, \ and\ \bibinfo
  {author} {\bibfnamefont {J.}~\bibnamefont {Luitz}},\ }\href@noop {}
  {\bibfield  {journal} {\bibinfo  {journal} {WIEN2k, An augmented plane
  wave+local orbitals program for calculating crystal properties}\ } (\bibinfo
  {year} {2001})}\BibitemShut {NoStop}%
\bibitem [{\citenamefont {Perdew}\ \emph {et~al.}(1996)\citenamefont {Perdew},
  \citenamefont {Burke},\ and\ \citenamefont {Ernzerhof}}]{pbe}%
  \BibitemOpen
  \bibfield  {author} {\bibinfo {author} {\bibfnamefont {J.~P.}\ \bibnamefont
  {Perdew}}, \bibinfo {author} {\bibfnamefont {K.}~\bibnamefont {Burke}}, \
  and\ \bibinfo {author} {\bibfnamefont {M.}~\bibnamefont {Ernzerhof}},\
  }\href@noop {} {\bibfield  {journal} {\bibinfo  {journal} {Phys. Rev. Lett.}\
  }\textbf {\bibinfo {volume} {77}},\ \bibinfo {pages} {3865} (\bibinfo {year}
  {1996})}\BibitemShut {NoStop}%
\bibitem [{\citenamefont {Dederichs}\ \emph {et~al.}(1984)\citenamefont
  {Dederichs}, \citenamefont {Blugel}, \citenamefont {Zeller},\ and\
  \citenamefont {Akai}}]{dederichs}%
  \BibitemOpen
  \bibfield  {author} {\bibinfo {author} {\bibfnamefont {P.~H.}\ \bibnamefont
  {Dederichs}}, \bibinfo {author} {\bibfnamefont {S.}~\bibnamefont {Blugel}},
  \bibinfo {author} {\bibfnamefont {R.}~\bibnamefont {Zeller}}, \ and\ \bibinfo
  {author} {\bibfnamefont {H.}~\bibnamefont {Akai}},\ }\href@noop {} {\bibfield
   {journal} {\bibinfo  {journal} {Phys. Rev. Lett.}\ }\textbf {\bibinfo
  {volume} {53}},\ \bibinfo {pages} {2512} (\bibinfo {year}
  {1984})}\BibitemShut {NoStop}%
\bibitem [{\citenamefont {Schwarz}\ and\ \citenamefont {Mohn}(1984)}]{fsm1}%
  \BibitemOpen
  \bibfield  {author} {\bibinfo {author} {\bibfnamefont {K.}~\bibnamefont
  {Schwarz}}\ and\ \bibinfo {author} {\bibfnamefont {P.}~\bibnamefont {Mohn}},\
  }\href@noop {} {\bibfield  {journal} {\bibinfo  {journal} {J. Phys. F}\
  }\textbf {\bibinfo {volume} {14}},\ \bibinfo {pages} {L129} (\bibinfo {year}
  {1984})}\BibitemShut {NoStop}%
\bibitem [{\citenamefont {Moruzzi}\ \emph {et~al.}(1986)\citenamefont
  {Moruzzi}, \citenamefont {Marcus}, \citenamefont {Schwarz},\ and\
  \citenamefont {Mohn}}]{fsm2}%
  \BibitemOpen
  \bibfield  {author} {\bibinfo {author} {\bibfnamefont {V.~L.}\ \bibnamefont
  {Moruzzi}}, \bibinfo {author} {\bibfnamefont {P.~M.}\ \bibnamefont {Marcus}},
  \bibinfo {author} {\bibfnamefont {K.}~\bibnamefont {Schwarz}}, \ and\
  \bibinfo {author} {\bibfnamefont {P.}~\bibnamefont {Mohn}},\ }\href@noop {}
  {\bibfield  {journal} {\bibinfo  {journal} {Phys. Rev. B}\ }\textbf {\bibinfo
  {volume} {34}},\ \bibinfo {pages} {1784} (\bibinfo {year}
  {1986})}\BibitemShut {NoStop}%
\bibitem [{\citenamefont {Lufaso}\ \emph {et~al.}(2006)\citenamefont {Lufaso},
  \citenamefont {Barnes},\ and\ \citenamefont {Woodward}}]{lufaso}%
  \BibitemOpen
  \bibfield  {author} {\bibinfo {author} {\bibfnamefont {M.~W.}\ \bibnamefont
  {Lufaso}}, \bibinfo {author} {\bibfnamefont {P.~W.}\ \bibnamefont {Barnes}},
  \ and\ \bibinfo {author} {\bibfnamefont {P.~M.}\ \bibnamefont {Woodward}},\
  }\href@noop {} {\bibfield  {journal} {\bibinfo  {journal} {Acta Cryst.}\
  }\textbf {\bibinfo {volume} {B62}},\ \bibinfo {pages} {397} (\bibinfo {year}
  {2006})}\BibitemShut {NoStop}%
\bibitem [{\citenamefont {Zhang}\ \emph {et~al.}(2018)\citenamefont {Zhang},
  \citenamefont {Reilly}, \citenamefont {Tkatchenko},\ and\ \citenamefont
  {Scheffler}}]{zhang}%
  \BibitemOpen
  \bibfield  {author} {\bibinfo {author} {\bibfnamefont {G.~X.}\ \bibnamefont
  {Zhang}}, \bibinfo {author} {\bibfnamefont {A.~M.}\ \bibnamefont {Reilly}},
  \bibinfo {author} {\bibfnamefont {A.}~\bibnamefont {Tkatchenko}}, \ and\
  \bibinfo {author} {\bibfnamefont {M.}~\bibnamefont {Scheffler}},\ }\href@noop
  {} {\bibfield  {journal} {\bibinfo  {journal} {New J. Phys.}\ }\textbf
  {\bibinfo {volume} {20}},\ \bibinfo {pages} {063020} (\bibinfo {year}
  {2018})}\BibitemShut {NoStop}%
\bibitem [{\citenamefont {Fu}\ and\ \citenamefont {Singh}(2019)}]{fu2}%
  \BibitemOpen
  \bibfield  {author} {\bibinfo {author} {\bibfnamefont {Y.}~\bibnamefont
  {Fu}}\ and\ \bibinfo {author} {\bibfnamefont {D.~J.}\ \bibnamefont {Singh}},\
  }\href@noop {} {\bibfield  {journal} {\bibinfo  {journal} {Phys. Rev. B}\
  }\textbf {\bibinfo {volume} {100}},\ \bibinfo {pages} {045126} (\bibinfo
  {year} {2019})}\BibitemShut {NoStop}%
\bibitem [{\citenamefont {Shimizu}(1981)}]{shimizu}%
  \BibitemOpen
  \bibfield  {author} {\bibinfo {author} {\bibfnamefont {M.}~\bibnamefont
  {Shimizu}},\ }\href@noop {} {\bibfield  {journal} {\bibinfo  {journal} {Rep.
  Prog. Phys.}\ }\textbf {\bibinfo {volume} {44}},\ \bibinfo {pages} {329}
  (\bibinfo {year} {1981})}\BibitemShut {NoStop}%
\bibitem [{\citenamefont {Aguayo}\ \emph {et~al.}(2004)\citenamefont {Aguayo},
  \citenamefont {Mazin},\ and\ \citenamefont {Singh}}]{aguayo}%
  \BibitemOpen
  \bibfield  {author} {\bibinfo {author} {\bibfnamefont {A.}~\bibnamefont
  {Aguayo}}, \bibinfo {author} {\bibfnamefont {I.~I.}\ \bibnamefont {Mazin}}, \
  and\ \bibinfo {author} {\bibfnamefont {D.~J.}\ \bibnamefont {Singh}},\
  }\href@noop {} {\bibfield  {journal} {\bibinfo  {journal} {Phys. Rev. Lett.}\
  }\textbf {\bibinfo {volume} {92}},\ \bibinfo {pages} {147201} (\bibinfo
  {year} {2004})}\BibitemShut {NoStop}%
\bibitem [{\citenamefont {Larson}\ \emph {et~al.}(2004)\citenamefont {Larson},
  \citenamefont {Mazin},\ and\ \citenamefont {Singh}}]{larson}%
  \BibitemOpen
  \bibfield  {author} {\bibinfo {author} {\bibfnamefont {P.}~\bibnamefont
  {Larson}}, \bibinfo {author} {\bibfnamefont {I.~I.}\ \bibnamefont {Mazin}}, \
  and\ \bibinfo {author} {\bibfnamefont {D.~J.}\ \bibnamefont {Singh}},\
  }\href@noop {} {\bibfield  {journal} {\bibinfo  {journal} {Phys. Rev. B}\
  }\textbf {\bibinfo {volume} {69}},\ \bibinfo {pages} {064429} (\bibinfo
  {year} {2004})}\BibitemShut {NoStop}%
\bibitem [{\citenamefont {Qin}\ \emph {et~al.}(2020)\citenamefont {Qin},
  \citenamefont {Ren},\ and\ \citenamefont {Singh}}]{Qin2020}%
  \BibitemOpen
  \bibfield  {author} {\bibinfo {author} {\bibfnamefont {G.}~\bibnamefont
  {Qin}}, \bibinfo {author} {\bibfnamefont {W.}~\bibnamefont {Ren}}, \ and\
  \bibinfo {author} {\bibfnamefont {D.~J.}\ \bibnamefont {Singh}},\ }\href@noop
  {} {\bibfield  {journal} {\bibinfo  {journal} {Phys. Rev. B}\ }\textbf
  {\bibinfo {volume} {101}},\ \bibinfo {pages} {014427} (\bibinfo {year}
  {2020})}\BibitemShut {NoStop}%
\end{thebibliography}%

\end{document}